\def\year{2016}\relax
\documentclass[letterpaper]{article}
\usepackage{aaai16}
\usepackage{times}
\usepackage{helvet}
\usepackage{courier}
\usepackage{multirow}
\usepackage{algorithm}
\usepackage{algpseudocode}
\usepackage{amsmath}
\usepackage{subfig}
\usepackage{enumitem}
\usepackage{algorithm}
\usepackage{algpseudocode}
\usepackage{amsmath}
\usepackage{subfig}
\usepackage{amssymb}
\usepackage{graphicx}
\usepackage{enumitem}
\usepackage{textcomp }
\usepackage[labelfont=bf]{caption}
\begin{document}
%
\title{Signed Link Analysis in Social Media Networks}
\author{
	Ghazaleh Beigi$^{\ast}$, Jiliang Tang$^{\dagger}$, and Huan Liu$^{\ast}$\\
	$^{\ast}${Arizona State University,}
	$^{\dagger}${Yahoo! Labs}\\
	$^{\ast}$\{gbeigi,huan.liu\}@asu.edu, $^{\dagger}$jlt@yahoo-inc.com
}

\maketitle
\begin{abstract}
Numerous real-world relations can be represented by signed networks with positive links (e.g., trust) and negative links (e.g., distrust). Link analysis plays a crucial role in understanding the link formation and can advance various tasks in social network analysis such as link prediction. The majority of existing works on link analysis have focused on unsigned social networks. The existence of negative links determines that properties and principles of signed networks are substantially distinct from those of unsigned networks, thus we need dedicated efforts on link analysis in signed social networks. In this paper, following social theories in link analysis in unsigned networks, we adopt three social science theories, namely \textit{Emotional Information}, \textit{Diffusion of Innovations} and \textit{Individual Personality}, to guide the task of link analysis in signed networks.
\end{abstract}
\section{Introduction}
The pervasive usage of social media allows users to participate in online activities that produce a large amount of social links. The task of link analysis aims to understand the factors influencing the link formation. The findings from link analysis have been exploited to help a variety of social media mining tasks such as trust prediction~\cite{gbeigi} and community detection~\cite{backstrom2006group}. 

Link analysis in unsigned social networks (or networks with only positive links) has been extensively studied. For example, users tend to create positive links to others who share certain similarity with them (Homophily), or two individuals geographically closer are more likely to become friends (Confounding). However, social networks in social media can contain both positive and negative links. Examples of such signed social networks include Epinions with trust/distrust links, and Slashdot with friend/foe links. The existence of negative links in signed networks challenges many existing concepts and properties of unsigned networks. For example, negative links present substantially distinct properties from positive links~\cite{szell2010multirelational} and Homophily of unsigned networks are not directly applicable to signed networks~\cite{tang2014distrust}. Therefore, link analysis for signed networks cannot be performed by simply extending that of unsigned networks, and hence requires further effort. 

There is a few recent works dedicated to link analysis in signed networks based on social theories (e.g., balance theory and status theory). On the one hand, these methods suggest that achievements from link analysis can benefit various signed network mining tasks such as link prediction; meanwhile they indicate that theories from social sciences can guide link analysis in signed networks. On the other hand, these methods merely focus on the topological structures, which could suffer from the link sparsity problem while completely ignore other pervasively available resources such as users' personalities and emotional information. Link analysis with leveraging these available resources not only greatly advance current research but also provide deep understandings of user behaviors in signed networks.  

The success of link analysis with social theories motivates us to adopt theories from psychology and social science to guide our link analysis in this paper, namely,  {\it Emotional Information}, {\it Diffusion of Innovations} and {\it Individual Personality}.  The first theory suggests that emotions of individuals toward each other are strong indicators of positive and negative links~\cite{bewsell2012distrust,dunn2005feeling}; the second one considers positive and negative link formation as a problem of an individual's probability of adopting a new behavior following her friends' behaviors~\cite{rogers2010diffusion}; and the last theory derived from~\cite{asendorpf1998personality}, suggests that people personality implies individuals' tendency to forming the positive and negative links.  Guided by these social theories, many interesting findings are revealed from our link analysis.
\section{Social Theories}
In this section, we explain the social theories \textit{Emotional Information}, \textit{Diffusion of Innovations} and \textit{Individual Personality}.
\subsubsection{Emotional Information Theory}
Users express their emotions toward each other in various ways. For example, in Slashdot users do so via commenting and replying to the posts; while product-review sites such as Epinions provide the rating mechanisms for their users to express their emotions. As a result, emotional information is pervasively available in social media no matter how they are exposed~\cite{beigi2016exploiting}. According to the sociologist, emotions of individuals toward each other, are strong indicators of positive and negative links~\cite{bewsell2012distrust,dunn2005feeling}. For example, happiness and satisfaction are indicators of positive emotions which could lead to the positive links; while negative emotions such as anger and fear imply the negative relations.
\subsubsection{Diffusion of Innovation Theory}
The problem of positive and negative link creation could relate to the well-studied topic of {\it diffusion of innovation}~\cite{rogers2010diffusion} by treating it as a behavior that spreads through the network. Therefore it turns into a new problem of analyzing an individual's tendency to follow her friends'  behaviors toward other users.
\subsubsection{Individuals Personality Theory}
According to~\cite{golbeck2011predicting}, people's behavior in social media could be a good indicator of their personality and the reasons are two-fold.  First, social media websites allow for free interaction and exposing viewpoints by providing an appropriate platform to satisfy users' basic psychological needs. Second, there is an ample amount of data regarding normative behaviors of individuals which guarantees fair analysis of individual's personality. Moreover, research from sociology suggest that people personality determines individuals' propensity to positive and negative relations~\cite{asendorpf1998personality}. For example, optimistic users are more grateful and receive more social support, and hence have higher chances in establishing and receiving positive links. In contrast, pessimists have negative attitudes, expect the worst of people, and treat positive events as flukes. Consequently, these individuals often receive or give negative links. Hence, considering user's personality information could be very helpful for studying the problem of signed link formation. Although there are many types of personalities~\cite{mccrae1998introduction}, in this paper we only consider two common ones, i.e., optimism and pessimism.
\section{Datasets}
We collect two large online signed social networks datasets from Epinions and Slashdot where individuals can express their opinions toward each other besides creating positive and negative links. Availability of product-rating data in Epinions and individuals' post reviews data in Slashdot, can help approximating user's personality in these websites.  For example optimistic users are likely to give higher rating scores in Epinions and likewise, more positive reviews on posts in Slashdot, whereas pessimistic users tend to give lower scores in Epinions and more negative comments in Slashdot~\cite{asendorpf1998personality}. Accordingly, we can define optimism and pessimism scores, based on the ratings in Epinions, and the users interactions in Slashdot.
\subsection{Epinions} Epinions is a product-review website where users can establish trust/distrust relationships toward each other. We treat each relation as either positive or negative links and construct user-user matrix ${\bf F}$ where ${\bf F}_{ij} = 1$ if user $i$ trusts user $j$, and ${\bf F}_{ij} = -1$ if user $i$ distrusts user $j$. Also, ${\bf F}_{ij} = 0$ where the information is missing. Users can express opinions toward each other by rating how helpful their reviews are, from 1 to 6. From these ratings, we construct the positive and negative emotion matrices ${\bf P}$ and ${\bf N}$ as follows: (1) we consider low helpfulness ratings $\{1,2\}$ as negative emotions, high helpfulness ratings $\{4,5,6\}$ as positive emotions and the rating $3$ as neutral and (2) for each pair of users $(u_i,u_j)$, we compute the number of positive and negative emotions expressed from $u_i$ to $u_j$ to create ${\bf P}_{ij}$ and ${\bf N}_{ij}$. 

We define the optimism and pessimism in Epinions as follows. Let $\mathcal{I} = \{I_1, I_2, \ldots, I_M\}$ be the set of $M$ items and assume $r_{ik}$ denotes the rating score from $u_i$ to $I_k$ with $r_{ik} = 0$ indicating that $u_i$ has not rated $I_k$ yet.  Also, consider $\overline{r}_k$ as the average rating score of the $k$-th item rated by users. In this paper, we consider scores in $\{1,2,3\}$ as low and $\{4,5\}$ as high scores. We use $\mathcal{OL}_i$ to denote the set of items with low average rating scores and rated by $u_i$:
\begin{align}
\mathcal{ OL}_i = \{I_k \mid r_{ki}\neq 0 \wedge \overline{r}_k\leq 3\} \nonumber 
\end{align}
We further use $\mathcal{OH}_i$ to denote the set of items which are scored high by $u_i$, and meanwhile have low average scores.  $\mathcal{OH}_i$ can be formally defined as:
\begin{align}
\mathcal{ OH}_i = \{I_k \mid I_k \in  \mathcal{ OL}_i \wedge {r}_{ik} > 3\} \nonumber 
\end{align}
Intuitively, the more frequently user $u_i$ has rated above the average, the more optimistic she is. Therefore we define the optimism score for $u_i$ as ${\bf o}_i = \frac{|\mathcal{ OH}_i|}{|\mathcal{ OL}_i|}$.  

Similarly we use $\mathcal{PH}_i$ to denote the set of items with high average rating scores and rated by $u_i$,
\begin{align}
\mathcal{PH}_i = \{I_k \mid r_{ki} \neq 0 \wedge \overline{r}_k > 3\} \nonumber 
\end{align}
Let $\mathcal{PL}_i$ denotes the subset of items from $\mathcal{PH}_i$, which are given low rates by $u_i$:
\begin{align}
\mathcal{ PL}_i = \{I_k \mid I_k \in  \mathcal{ PH}_i \wedge {r}_{ik} \leq 3\} \nonumber 
\end{align}
We define the pessimism score $u_i$ as: ${\bf p}_i = \frac{|\mathcal{ PL}_i|}{|\mathcal{ PH}_i|}$.

\subsection{Slashdot} Slashdot is a technology-related news platform which allows users to tag each other as either `friend' or `foe'. Similar to the Epinions, we construct user-user matrix ${\bf F}$ from the positive and negative links in the network. Likewise, users can express their opinions and comments toward each other by annotating the articles posted by each other. In a similar way to the Epinions, using positive and negative opinions, we create user-user positive and negative emotion matrices ${\bf P}$ and ${\bf N}$ by computing the number of positive or negative emotions users express toward each other.

Likewise, we define individual's personality in Slashdot based on user-user emotion matrices ${\bf P}$ and ${\bf N}$. Let $\overline{P}$ and $\overline{N}$ be the average of positive and negative emotions between all pairs of users, respectively. We also define $\overline{{\bf P}}_j$ and $\overline{{\bf N}}_j$ as the average of positive and negative emotions received by $u_j$. Further, we define $\mathcal{OL}_i$, as a set of users $u_j$ who have received positive emotions from $u_i$, but at the same time, have received more negative emotions than the average in the network, i.e. they are worser than the average,
\begin{align}
\mathcal{ OL}_i = \{u_j \mid {\bf P}_{ij}\neq 0 \wedge \overline{{\bf N}}_j> \overline{N}\} \nonumber 
\end{align}
We formally define $\mathcal{OH}_i$ to denote the set of users $u_k$ who belong to $\mathcal{ OL}_i$ and have received more positive emotions from $u_i$ than $\overline{{\bf P}}_k$,
\begin{align}
\mathcal{ OH}_i = \{u_k \mid u_k \in  \mathcal{ OL}_i \wedge {\bf P}_{ik} > \overline{{\bf P}}_k\} \nonumber 
\end{align}
In other words, the more frequently $u_i$ has given positive emotions to the worst users in the network, the more optimistic she is. Therefore we define the optimism score for $u_i$ as ${\bf o}_i = \frac{|\mathcal{ OH}_i|}{|\mathcal{ OL}_i|}$.

Likewise, we define $\mathcal{PH}_i$, as a set of users $u_j$ who have received negative emotions from $u_i$, but at the same time, have received more positive emotions than the average in the network, i.e. they are better than the average,
\begin{align}
\mathcal{PH}_i = \{u_j \mid {\bf N}_{ij} \neq 0 \wedge \overline{{\bf P}}_j > \overline{P}\} \nonumber 
\end{align}
We define $\mathcal{PL}_i$ to denote the set of users $u_k$ who belong to $\mathcal{ PH}_i$ and have received more negative emotions from $u_i$ than $\overline{{\bf N}}_k$,
\begin{align}
\mathcal{ PL}_i = \{u_k \mid u_k \in  \mathcal{ PH}_i \wedge {{\bf N}}_{ik} > \overline{N}_k\} \nonumber 
\end{align}
Pessimism of $u_i$ could be similarly defined as: ${\bf p}_i = \frac{|\mathcal{ PL}_i|}{|\mathcal{ PH}_i|}$.

\noindent There might be other ways to construct ${\bf P}$, ${\bf N}$, ${\bf o}$ and ${\bf p}$ from the data which we leave to future work. Table \ref{Tab:DataStat} shows the statistics of our datasets.

\begin{table}[!htbp]
	\centering
	\small
	\caption{\textbf{Statistics of the Raw Data.}}\label{Tab:DataStat}
	\begin{tabular}{l|l|l} 
		& Epinions & Slashdot  \\ \hline \hline
		\# of Users & 405,178 & 7,275\\
		\# of Positive Links & 717,677 & 67,705 \\
		\# of Negative Links & 123,705 & 20851 \\		
		\# of Positive Emotions & 12,581,748 & 1,742,763\\
		\# of Negative Emotions & 319,908 & 42,260\\ \hline
	\end{tabular}
\end{table}
\section{Data Analysis and Observations}
In this section, we investigate how each social theory is related to the formation of positive and negative links. 
\subsection{Emotional Information Theory}
 Here, we investigate (1) the existence of the correlation between emotional information and positive/negative links in signed social networks and, (2) study the impact of emotional strength on the formation of positive and negative relations. Specifically, we aim to answer the two questions:
\begin{itemize}
	\item Are users with positive (negative) emotions more likely to establish positive (negative) relations than those without?
	\item Are users with higher positive (negative) emotion strengths more likely to create positive (negative) links than those with lower positive (negative) emotion strengths?
\end{itemize}

To answer the {\bf first question}, for each pair of users $(u_i, u_j)$ with positive emotions, we randomly select a user $u_k$ where there is no positive emotions from $u_i$ to $u_k$. We then check the existence of positive relations from $u_i$ to $u_j$ and $u_i$ to $u_k$, respectively. We set $vp = 1$ if ${\bf F}_{ij} = 1$ and $vp = 0$ otherwise. Likewise, we set $vr = 1$ if ${\bf F}_{ik}=1$ and $vr = 0$ otherwise. This way, we obtain two vectors, ${\bf v}_p$ and ${\bf v}_r$ where ${\bf v}_p$ is the set of all $vp$s for pairs of users with positive emotions and ${\bf v}_r$ is the set of $vr$s for pairs of users without positive emotions. We conduct a two-sample $t$-test on ${\bf v}_p$ and ${\bf v}_r$ as follows:
\begin{align}
H_0: {\bf v}_p \leq {\bf v}_r,~~~H_1: {\bf v}_p > {\bf v}_r
\end{align}
\noindent where the null hypothesis is rejected at significance level $\alpha = 0.01$ with p-values of $4.32e{-62}$ and $6.17e{-47}$ over Epinions and Slashdot, respectively. A similar procedure can be followed for negative emotions which we omit the details for brevity. Results from $t$-test suggest that {\it with high probability, users with positive (negative) emotions are more likely to establish positive (negative) links than those without.}

To answer the {\bf second question}, we rank all pairs of users $(u_i, u_j)$ with positive emotions according to their emotion strengths (${\bf P}_{ij}$) in a descending order and divide those pairs into $K$ groups $\mathcal{E} = \{\mathcal{E}_1,\mathcal{E}_2,\ldots,\mathcal{E}_K\}$ with equal sizes. The emotion strengths in $\mathcal{E}_i$ are larger than those in $\mathcal{E}_j$ if $i < j$. Then we form $\frac{K(K-1)}{2}$ pairs of groups $(\mathcal{E}_i,\mathcal{E}_j)$ with $i < j$ where $\mathcal{E}_i$ is the group with higher emotional strengths and $\mathcal{E}_j$ is the one with lower emotional strengths. For each pair of groups, we use $hp$ and $lp$ to denote the number of pairs of users with positive relations in $\mathcal{E}_i$ and $\mathcal{E}_j$, receptively. By repeating this over all pairs of groups, we can obtain two vectors ${\bf h}$ and ${\bf l}$ for $hp$s and $lp$s, respectively.

We conduct a two-sample $t$-test on ${\bf h}$ and ${\bf l}$ by defining the null hypothesis $H_0$ as users with weak positive emotion strengths are more likely to establish positive links and the alternative hypothesis $H_1$ as users with strong positive emotion strengths are more likely to create positive:
\begin{align}
H_0: {\bf h} \leq {\bf l} ,~~~ H_1: {\bf h} > {\bf l}.
\end{align}
By choosing $K = 10$, the null hypothesis is rejected at significance level $0.01$ with p-values of $8.47{e-23}$ and $1.72{e-19}$ for Epinions and Slashdot, respectively . We make similar observations when we set $K = 30$ and $K = 50$. Similarly, we observe the impact of the strengths of negative emotions on the formation of negative links. These results suggest that \textit{users with higher positive (negative) emotion strengths are more likely to establish positive (negative) links than those with lower positive emotion strengths}.

\subsection{Diffusion of Innovation Theory}
 Following the diffusion of innovation theory, our goal here is to study if the behavior of user $u_i$ toward user $u_j$ could be influenced by the behavior of $u_i$'s friend $u_k$ toward $u_j$. More specifically, we aim to answer the following question:
\begin{itemize}
	\item Is user $u_i$ with a friend $u_k$ who has a positive (negative) link to user $u_j$, more likely to establish a positive (negative) link with $u_j$ than if he/she does not have such friend?
\end{itemize}
To answer this question, we first find a pair of users $(u_i, u_j)$  where $u_i$'s friend $u_k$ has a positive link to $u_j$. We also randomly select a user $u_r$ without any positive relations with $u_k$. We then check if there are positive links from $u_i$ to $u_j$ and from $u_i$ to $u_r$. We set $fp = 1$ if ${\bf F}_{ij}=1$ and $fp = 0$ otherwise; Similarly, we set $fr = 1$ if ${\bf F}_{ir}=1$ and $fr = 0$ otherwise. We then construct two vectors, ${\bf f}_p$ and ${\bf f}_r$ where ${\bf f}_p$ is the set of all $fp$s and ${\bf f}_r$ is the set of $fr$s.

\begin{figure}[ht]
	\centering
	{\includegraphics[width=0.25\textwidth]{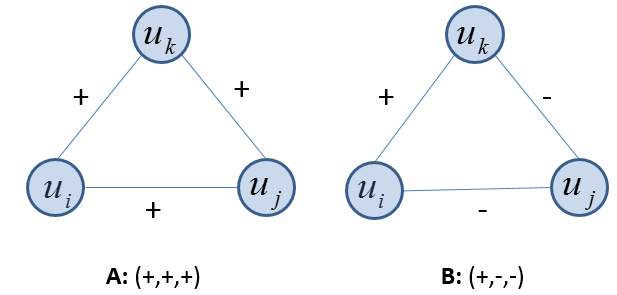}}
	\caption{\textbf{Balanced Triads Encouraged by Diffusion of Innovation Theory.}}\label{fig:BalanceTriads}
\end{figure}

We conduct a two-sample $t$-test on ${\bf f}_p$ and ${\bf f}_r$ with the null and alternative hypotheses $H_0$ and $H_1$ defined as follows:
\begin{align}
H_0: {\bf f}_p \leq {\bf f}_r,~~~H_1: {\bf f}_p > {\bf f}_r
\end{align}
\noindent where the null hypothesis is rejected at significance level $\alpha = 0.01$ with p-values of $1.84e{-75}$ and $3.56e{-91}$ over Epinions and Slashdot, respectively. Likewise, we repeat the process for friends with negative links; however for brevity we omit the details and directly give the suggestions from the results of the two-sample $t$-test as follows: {\it users are likely to follow their friends' behaviors in terms of positive and negative link creation.}  The theory is likely to encourage triads shown in Figure~\ref{fig:BalanceTriads}, which are balanced according to balance theory.

\subsection{Individuals Personality Theory}\label{section:IP}
Next, we investigate the impact of user's personality on the formation of positive and negative links via studying the correlation between personality information and positive and negative links. We seek to answer the following questions:
\begin{itemize}
	\item Are users with higher optimism, more likely to establish positive links than those with lower optimism? and 
	\item Are users with higher pessimism, more likely to create negative links than those with lower pessimism?
\end{itemize}

To answer the {\bf first question}, we rank all users in a descending order according to their optimism scores and divide them into $K$ levels with equal sizes denoted as $G=\{g_1, g_2,..., g_K\}$. There are $\frac{K(K-1)}{2}$ pairs of $(g_i,g_j)$ where $i<j$. We consider $g_i$ as the group of more optimistic users compared to those in $g_j$. For each pair $(g_i,g_j)$, we use $H$ and $L$ to denote the number of positive links established by users in groups $g_i$ and $g_j$, respectively. Therefore, we have two vectors ${\bf h}$ and ${\bf l}$ for $H$s and $L$s of all pairs of groups.

We conduct a two-sample $t$-test on ${\bf h}$ and ${\bf l}$ where $H_0$ is that users who are less optimistic are more likely to establish positive links and $H_1$ is that users with higher level of optimism are more likely to create positive relations:
\begin{align}
H_0: {\bf h} \leq {\bf l} ,~~~ H_1: {\bf h} > {\bf l}.
\end{align}
We set $K = 20$, and the null hypothesis is rejected at significance level $0.01$ with p-values $3.16{e-19}$ and $1.6029{e-23}$ for Epinions and Slashdot datasets, respectively. We make similar observations by setting $K = 30$ and $K = 50$. These results suggest that \textit{users with high optimistic behavior are more likely to establish positive links than those with low optimism}. Following a similar procedure, we observe that \textit{users who are more pessimistic are more likely to establish negative relations than those with low level of pessimism}.

\section{Conclusion and Future Work}
In this study, we employ three theories from psychology and social sciences, namely \textit{Emotional Information}, \textit{Diffusion of Innovations} and \textit{Individual Personality}, on the additional information available in the networks including user's emotional and personality, for the task of link analysis in signed networks. In future, we plan to replicate this study by exploiting other available information in the network. Another interesting research direction is to study link analysis in dynamic signed networks by deploying the social theories. 
\section{Acknowledgments}
This material is based upon the work supported by, or in part by, Army Research Office (ARO) under grant numbers W911NF-15-1-0328 and \#025071.
\bibliographystyle{aaai}
\bibliography{formatting-instructions-latex}

\end{document}